\documentclass[a4paper]{mem}
\usepackage{natbib}
\usepackage{graphicx}
\usepackage[a4paper]{hyperref}
\idline{73}{1}
\begin{document}
   \title{The EPIC/MOS view of the 2-8 keV Cosmic X-ray Background 
Spectrum}

\author{Andrea De Luca\inst{1,2} \and Silvano Molendi\inst{1}}

   \offprints{A. De Luca}
\mail{via Bassini 15, 20133 Milano}

   \institute{INAF/IASF ``G. Occhialini'', Via Bassini 15, I-20133 
Milano, Italy \email{deluca@mi.iasf.cnr.it}
              \and  Universit\`a di Milano Bicocca, Dipartimento di 
Fisica, P.za della Scienza 4, I-20126 Milano, Italy
             }

   \abstract{We have measured the spectrum of the Cosmic X-ray Background (CXB) 
in the 2-8 keV range with the high throughput EPIC/MOS instrument onboard 
XMM-Newton. A large sample of high galactic latitude observations was used, 
covering a total solid angle of 5.5 square degrees. Our study is based on a 
very 
careful characterization and subtraction of the instrumental background, which 
is crucial for a robust measurement of the faintest diffuse source of the 
X-ray 
sky. The CXB spectrum is consistent with a power law having a photon index 
$\Gamma=1.41\pm0.06$, with a 2-10 keV flux of 
(2.24$\pm$0.16)$\times10^{-11}$ erg cm$^{-2}$ s$^{-1}$ deg$^{-1}$
(90\% confidence level, including the absolute flux calibration uncertainty). 
Our results 
are in excellent agreement with two of the most recent CXB measurements, 
performed with BeppoSAX  LECS/MECS data (Vecchi et al. 1999) and with an 
independent analysis of XMM-Newton EPIC/MOS data (Lumb et al. 2002), 
providing a very strong constrain to the absolute sky surface brightness in 
this 
energy range, so far affected by a $\sim$40\% uncertainty. Our measurement 
immediately implies that the fraction of CXB resolved by the recent deep X-ray 
observations in the 2-10 keV band is of $80\pm7$\% (1$\sigma$), suggesting the existence 
of 
a new population of faint sources, largely undetected within the current 
sensitivity limits of the deepest X-ray surveys.

   }
   \authorrunning{A. De Luca \& S. Molendi}
   \titlerunning{EPIC Cosmic X-ray Background Spectrum}
   \maketitle

\section{Introduction}

\label{intro}

The discovery of a diffuse background radiation in the X-ray sky 
dates back
to the birth of X-ray astronomy: the first evidence was obtained by
during the same rocket experiment which led to the
discovery of Sco X-1, the first extra-solar X-ray source (Giacconi 
et al. 1962).
Later observations have demonstrated that the bulk of Cosmic X-ray 
Background (CXB)
above energies of $\approx$2 keV is of extragalactic origin, due to 
sources
below the detection threshold.
The first wide band measures of the CXB were 
made by HEAO$-$1 (1977): the CXB spectrum in the 2$\div$10 keV 
range was 
well described by a simple power law with photon index $\approx$ 
1.4 (Marshall et al. 1980).
More recently, several measurements of the CXB spectrum 
have been obtained at energies below 10 keV. While the results on 
the spectral 
shape confirmed a power law with $\Gamma\sim1.4$, the normalization 
of the CXB 
remained highly uncertain as a consequence of large discrepancies 
(well beyond 
the statistical errors) among the different determinations. A 
difference as 
large as $\sim40$\% is found from the highest measured value 
(Vecchi et al. 1999, using SAX data) to the lowest one (the 
original HEAO-1 experiment, Marshall et al. 1980).\\
Barcons et al. (2000)
showed that cosmic variance cannot account for the differences 
among the 
previous measures of the CXB intensity, concluding that systematic 
errors and cross-calibration differences must have some 
role. Measurements published after the analysis of  Barcons et al. 
(2000), namely Lumb et al. (2002) with XMM-Newton EPIC and 
Kushino et 
al. (2002) with ASCA/GIS, while differing by $\sim$15\% only, 
either because of 
the small covered solid angle (Lumb et al. 2002), or because of 
large 
uncertainties in 
the stray light assessment (Kushino et al. 2002), do not allow to 
constrain the 
value 
of the CXB normalization to a much narrower range.\\
Such an uncertainty on the intensity represents a severe limit 
to the definition of a big picture, explaining which are the 
sources of the CXB and 
constraining their cosmological properties, in spite of the wealth 
of information coming from the deep pencil-beam 
X-ray surveys (Chandra Deep Field North, Brandt et al. 2001; 
Chandra Deep Field 
South, Giacconi et al. 2002), medium-deep wide angle X-ray surveys 
(HELLAS2XMM, 
Baldi et al. 2002) and their multiwavelength follow-up
Campaigns. Even a basic information such 
as the resolved fraction of the CXB cannot be firmly evaluated.   \\
We have obtained a new measurement of the CXB spectrum in the 2-8 
keV range with the EPIC/MOS\footnote{ the {\em pn} camera,
having different characteristics, will require a different 
approach} cameras onboard XMM-Newton. 
Our study of the CXB is based on a large sample of high galactic latitude 
pointings for a total 
solid angle of $\sim$5.5 square degrees, reducing the effects of 
cosmic variance. Our analysis includes a very robust characterization 
of the instrumental background properties, which is crucial for a 
robust measurement of the faintest diffuse source of the 
X-ray sky. Here we shall briefly report on some highlights of our 
study. The reader is referred to De Luca \& Molendi (2004) for more 
details.

\section{The EPIC/MOS instrumental background}

The EPIC/MOS cameras (Turner et al., 2001) have
appropriate characteristics to study extended sources with low 
surface
brighteness, offering a large collecting area ($\approx $ 
cm$^2$) and good spectral resolution (2\% @ 6 keV) over a broad 
energy 
range (0.2$\div$12 keV) and a wide field of view ($\approx$ 15 
arcmin radius).
However such detectors suffer a rather high instrumental background 
(Non X-ray 
Background, NXB). 

\subsection{Different components of the MOS NXB}

The EPIC NXB can be divided into two parts: an electronic noise 
component, 
which is important only at the lower energies (below $\approx$0.3 
keV), and  a 
particle-induced component which dominates above 0.3 keV 
and is due to the interaction of particles in the orbital 
environment with the 
detectors and the structures that surround them. 
The particle-induced NXB is the sum of two different components:
\begin{itemize}
\item Low energy particles (E$\sim$ a few tens of keV) accelerated 
in the 
Earth magnetosphere can reach the detectors scattering 
through 
the telescope mirrors, generating
events which are almost indistinguishable from valid X-ray photons. 
When a concentrated cloud of 
such particles is channeled by the telescope mirrors to the focal 
plane, a sudden increase (up to three orders of magnitude)
of the quiescent count rate is observed (the so-called ``soft proton flares'')
An extreme time variability (time scale ranging from $\sim100$ s to
several hours) is the 
fingerprint of this background component; it will be hereafter 
called the ``flaring'' NXB or ``Soft Proton'' (SP) NXB.\\
An additional component of background can be generated by a steady 
flux of low energy particles, reaching the detectors through the 
telescope optics at a uniform rate. 
\item High energy particles (E$>$ a few MeV) generate a signal 
which is mostly 
discarded on-board on the basis of an upper energy thresholding and 
of 
a PATTERN analysis of the events (see e.g. Lumb et al. 2002). The 
unrejected 
part of this signal represent an important component of the MOS 
NXB. 
Its temporal behaviour is driven by the flux of energetic 
particles; 
its variability has therefore a time scale much larger than the 
length 
of a typical observation. We will refer to this NXB component as to 
the ``quiescent'' background.
\end{itemize}

\subsection{How to deal with the MOS NXB}
It is quite easy to identify the flaring background: a light curve can 
immediately show the time 
intervals affected by an high background count rate. Such intervals 
are unusable for the analysis of faint diffuse sources and have to 
be 
rejected with the so-called Good Time Interval (GTI) filtering.\\
After the application of the GTI, a residual component of soft 
proton  
background may survive. Low-amplitude flares, yielding 
little 
variations to the quiescent count rate, could be missed by the GTI 
threshold. Moreover, a slow time variability could hamper the 
identification of a ``flare'' by means of a 
time 
variability analysis. In such cases the unrejected NXB component 
may 
be revealed with a surface brightness analysis. As stated before, 
low 
energy particles are focused by the mirrors and therefore the 
spatial 
distribution of the induced NXB varies across the plane of the 
detectors. 
A rather large portion of the MOS detectors is not exposed to the sky 
(hereafter called ``out Field Of View'', {\em out FOV}) and 
therefore 
neither cosmic X-ray photons nor low energy particle induced events 
are 
collected there.
A comparison of the {\em out FOV} and {\em in FOV} surface brightness 
allows to identify and reject the 
observations affected by an anomalous low-energy particle NXB.\\
The final step required to remove the effects of NXB is to account 
for 
the quiescent component. The only way to solve the problem is 
to get an independent 
measurement of its spectrum. 
Its subtraction from the total (CXB$+$quiescent 
NXB) spectrum yields the pure 
CXB spectrum. The crucial problem is that the NXB spectrum, 
resulting 
from a non-symultaneous measurement, must be  representative of the 
actual 
quiescent NXB which is present in a typical observation of the sky.
There are two ways to measure the quiescent NXB in the MOS. 
Firstly, 
through the 
analysis of the {\em out FOV} regions, where neither X-ray photons nor 
soft 
protons  
can reach the focal plane through reflections/scatterings by the 
telescope 
optics. Secondly, through the study of the observations with 
the filter wheel in {\em closed} position:  an 
aluminium window prevents X-ray photons and low energy particles 
from 
reaching the detectors. 
Our analysis (see De Luca \& Molendi 2004 for details) fully confirmed 
the feasibility of such approach, favouring  the 
choice of the {\em closed} observations over the {\em out FOV}, as the former
provides an NXB spectrum better suited to study the CXB.

\section{Data preparation and analysis}
Our study is based on a rather large sample of MOS data including 
calibration, performance verification and granted time 
observations; 
public observations retrieved through the XMM-Newton Science 
Archive 
facility were also used. The initial dataset consist of a 
compilation 
of (mostly) blank sky fields observations and a list of 
observations 
performed with the filter wheel in {\em closed} position.
We selected only high galactic latitude fields 
($|b|\ge28^{\circ}$). We avoided pointings towards the Magellanic Clouds, 
Cluster of Galaxies, as well as observations of very bright 
targets. The selected fields were observed between revolution number 57 
and revolution number 437. We retrieved the {\em closed} observations 
performed 
in the same time interval, between revolution 25 and 462.
In our analysis we included all of the serendipitous sources detected
in the sky fields. We cut out only the bright target (if any) of the
observation in order to get an unbiased measure of the CXB.\\
We have developed an ad-hoc 
pipeline to perform the different steps of the analysis,
from raw data to the extraction of the spectra, in an 
automated way. 
A detailed description of such algorithm is given by De Luca \& Molendi 
(2004) and will not be reported here.

\section{Spectral analysis and results}

   \begin{figure}
   \centering
   \includegraphics[angle=-90,width=6cm]{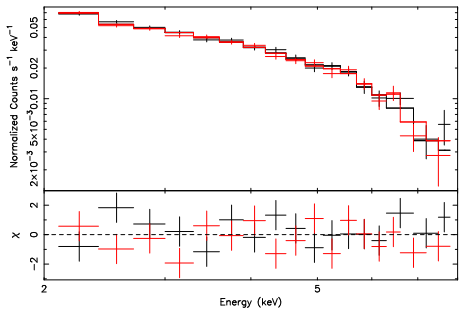}
      \caption{The cosmic X-ray background spectrum in the 2-8 keV 
range is displayed, folded with the instrumental response. MOS1 data are 
represented in red, MOS2 in black. The best fit model is overplotted. The 
lower 
panel shows the residuals in units of sigma.}
   \end{figure}

The final dataset includes 42 sky fields for the MOS1 camera and 43 for the 
MOS2. The total exposure time is of $\sim$1.15 Ms per camera. The solid angle 
covered by the data, summing the contribution of each observation (accounting 
for the differences in field of view due to the readout mode or to the 
excision 
of the central target) is of $\sim$5.5 square degrees (34 different pointing 
directions) per camera.
The {\em closed} data amount to $\sim$430 ksec per camera.\\
The 2-8 keV range was selected for the analysis. Lower energies were not used 
to 
avoid (i) contaminations by the soft galactic component of the CXB (emerging 
below $\sim$1 keV) and (ii) possible artefacts due to an imperfect subtraction 
of the bright internal Al-K and Si-K fluorescence lines in the 1-2 keV range. 
Above 8 keV the collected CXB signal is marginal.\\
The spectral analysis was performed within XSPEC v11.0.
The spectral model was a simple absorbed power law. The interstellar 
absorption 
$N_H$ was fixed to the exposure-weighted average of the values of the 
selected fields.\\
The spectrum of the cosmic X-ray background as seen by the MOS instruments is 
shown in Fig.~1. 
The two cameras yield fully consistent results within the statistical 
uncertainties. A symultaneous fit to the data 
($\chi^{2}_{\nu}=1.15$, 72 d.o.f.) yields a photon index of 1.41$\pm$0.04 and 
a 
normalization of $2.647\pm0.038$ photons cm$^{-2}$ s$^{-1}$ sr$^{-1}$ 
keV$^{-1}$ 
at 3 keV (to be corrected for the stray light, i.e. the contribution to the 
collected flux due to photons coming from out-of-field angles). The quoted 
uncertainties 
are the statistical errors at the 90\% confidence level for a single 
interesting parameter. \\
A careful study of the possible sources of errors (see De Luca \& Molendi, 2004) led us to compute the overall 
uncertainty (systematics included) to be of 4\% for the photon index and of 
3.5\% for the normalization. \\
After correcting for the stray light (7$\pm$2\% of the collected flux, according to Lumb et al. 2002), the MOS results on the 2-8 keV CXB spectrum are:
$$\Gamma\,=\,1.41\pm0.06$$
%\vspace{-1cm}
$$N\,=\,2.462\pm0.086$$ 
where the normalization $N$ is expressed in photons cm$^{-2}$ s$^{-1}$ sr$^{-1}$ keV$^{-1}$ at 3 keV.\\
The resulting flux in the 2-10 keV energy range is of 
(2.24$\pm$0.16)$\times10^{-11}$ erg   
cm$^{-2}$ s$^{-1}$ deg$^{-2}$. 
 The error (90\% confidence) includes also an extra 5\% uncertainty as an 
estimate
of the absolute flux calibration accuracy of the MOS cameras.
To ease a comparison with previous works, the 
corresponding normalization at 1 keV is of $\sim$11.6 photons cm$^{-2}$ 
s$^{-1}$ 
sr$^{-1}$ keV$^{-1}$.

\section{Discussion}

  \begin{figure*}
   \centering
  \includegraphics[angle=-90,width=9.5cm]{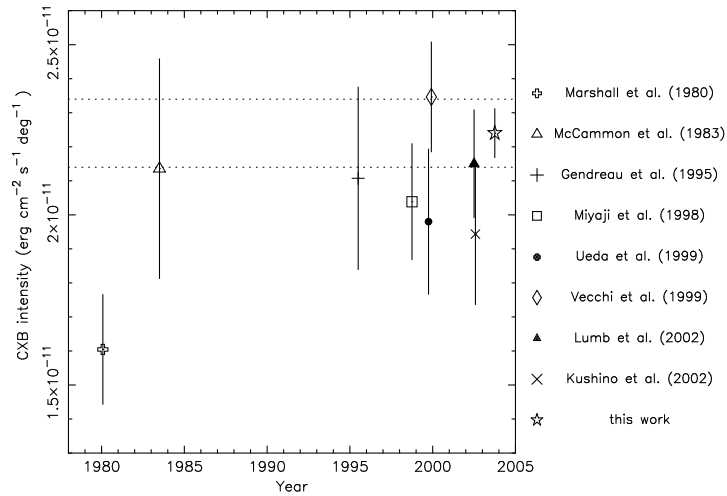}
   \caption{CXB intensity measurements. The flux in the 2-10 keV band is 
represented 
 as a function of the epoch of the experiment. The plot is an update of Fig.3 of 
Moretti et al. (2003) including the results of the present work.  From left to 
right the CXB values are from Marshall et al. (1980) with HEAO-1 data; McCammon 
et al. 
(1983) with a rocket measurement; Gendreau et al. (1995) with ASCA SIS data; 
Miyaji et al. (1998) with ASCA GIS; 
Ueda et al. (1999) with ASCA GIS and SIS; Vecchi et al. (1999) with BeppoSAX 
LECS/MECS; Lumb et al. (2002) with XMM-Newton EPIC/MOS; Kushino et al. (2002) 
with 
ASCA GIS. Finally, the hollow star mark our own CXB measurement with the EPIC 
MOS 
cameras onboard XMM-Newton. All the uncertainties are at 1$\sigma$ level. 
The horizontal dotted lines mark the range of CXB intensity constrained by our 
measurement after including an extra 3.2\% uncertainty (1$\sigma$ confidence
 level, rescaled from the 5\% uncertainty quoted in the text, which corresponds
to 90\% confidence level) on the absolute flux 
calibration of the EPIC cameras (it is indeed a measure of the cross-calibration 
accuracy among different instruments). Such range represents the most robust
estimate of the true absolute sky surface brightness in the 2-8 keV band.
}
    \end{figure*}

We have plotted in Figure 2 our measurement of the CXB intensity,
together with previous determinations. Our value is fully compatible with
the results of Vecchi et al. (1999) on BeppoSAX LECS/MECS data (total
solid angle $\Omega \sim 0.77$ square degrees) and with the independent
analysis of XMM-Newton EPIC/MOS data by Lumb et al. (2002) ($\Omega \sim1.2$ 
square degrees).\\
The ASCA/GIS measure of Kushino et al. (2002) is marginally consistent
with ours. Such study was performed over a rather large solid
angle ($\Omega \sim50$ square degrees) 
but the absolute flux determination accuracy was 
limited to 10\% by the large stray light component 
($\sim40$\% of the collected flux) affecting ASCA data.
Very recently, Revnivtsev et al. (2003) published a new measurement
of the CXB spectrum performed with the PCA instrument onboard RXTE,
reporting an intensity consistent with the results of Kushino et al. (2002).
They used data from nearly all the sky ($\Omega \sim 2.2\,10^4$ square 
degrees); the instrumental background spectrum was extracted from only 
25 ksec of dark Earth observation.
The original HEAO-1 measurement (Marshall et al., 1980) is significantly
lower than ours. Such measurement is very robust as for solid angle
coverage ($\Omega \sim10^4$ square degrees); however, all of the subsequent
determinations yielded invariably higher values of the CXB intensity,
casting some doubts on the absolute flux calibration of the HEAO-1 
instruments.\\
In our analysis, we used a compilation of sky fields covering a solid angle
of $\sim5.5$ square degrees, reducing the cosmic variance below the overall
quoted uncertainty. Our measurement, performed with the well-calibrated
EPIC/MOS instrument, relies on a very careful study and subtraction of the 
NXB, as well as on a detailed analysis of all the possible sources of errors.
In conclusion, we believe that our measure, fully consistent with two
of the most recent CXB determinations, set a very strong constrain on the 
CXB intensity in the 2-8 keV range, significantly higher than the former result from 
HEAO-1 data, assumed more than 20 years ago as a reference.\\
We can now compare our value of the CXB intansity with the source number 
counts derived from recent X-ray surveys, both deep/pencil-beam and
medium-deep/wide-angle (Moretti et al., 2003). 
Our measurement of the CXB intensity, 
F$_{CXB}$=(2.24$\pm$0.10)$\times10^{-11}$ erg cm$^{-2}$ s$^{-1}$ deg$^{-1}$
 ($1\sigma$ error, including the absolute flux uncertainty), 
implies that 80$^{+7}_{-6}$\% of the cosmic X-ray background has been 
resolved into discrete sources in the 2-10 keV band. The resolved fraction
rises only to $\sim$84\% when extrapolating the LogN/LogS  down to fluxes of $\sim10^{-17}$ erg cm$^{-2}$ s$^{-1}$, a factor of 10
below the sensitivity limits of the deepest X-ray surveys.
Such a result suggests the existence of a new class of X-ray sources 
(possibly heavily absorbed AGNs, or star-forming galaxies, see Moretti 
et al. 2003 and references therein) emerging at fluxes fainter than $10^{-16}$ 
erg cm$^{-2}$ s$^{-1}$ and accounting for the 
remaining part of unresolved CXB.

\bibliographystyle{aa}

\end{document}